\documentclass[aps,showpacs,natbib,eqsecnum]{revtex4}

\usepackage{graphicx,color}
\usepackage{amssymb}

\def\ben{\begin{equation}}
\def\ene{\end{equation}}
\mathchardef\bigtilde="0365

\def\la{\langle}
\def\ra{\rangle}
\def\s{\sigma}

\begin{document}

\title{%
Delta Expansion on the Lattice and Dilated Scaling Region
}

\author{Hirofumi Yamada}\email{yamada.hirofumi@it-chiba.ac.jp}
\affiliation{%
Division of Mathematics, Chiba Institute of Technology, Shibazono 2-1-1, Narashino, Chiba 275-0023, Japan}

\date{\today}
\begin{abstract}%
A new kind of delta expansion is applied on the lattice to the $d=2$ non-linear $\sigma$ model at $N=\infty$ and $N=1$ which corresponds to the Ising model.  We introduce the parameter $\delta$ for the dilation of the scaling region of the model with the replacement of the lattice spacing $a$ to $(1-\delta)^{1/2}a$.  Then, we demonstrate that the expansion in $\delta$ admits an approximation of the scaling behavior of the model at both limits of $N$ from the information at a large lattice spacing $a$.
\end{abstract}


\maketitle

\section{Introduction}
It is one of the outstanding problems to construct a systematic computational framework to study nonperturbative aspects of quantum fields.  Lattice field theories initiated by Wilson \cite{wil} allow us to use the strong coupling expansion which is mathematically equivalent with the high temperature expansion in condensed matter physics.  However, the results do not necessarily provide us, at least in the quantitative sense, the corresponding results in continuum space-time since the strong coupling expansion on the lattice usually breaks down at small lattice spacings.  Nevertheless the fact that the strong coupling expansion on the lattice clarified various nonperturbative properties of quantum fields such as the quark confinement serves us enough motivation to investigate the possibilities of improving it as to be effective on the approximation of physics in the continuum limit. 

As an attempt toward the improvement of the strong coupling expansion on the lattice, we propose a new computational scheme which may be considered as an alternative to the ordinary delta expansion \cite{early}\cite{dun}.   We introduce $\delta$ as the parameter to dilate the scaling region of a given lattice model with the replacement of the lattice spacing $a$ to $(1-\delta)^{1/2}a$.  As long as $\delta$ can be tuned to some values close to unity, the lattice spacing $a$ may be kept large enough in calculating physical quantities near the continuum limit.  Further we perform the expansion in $\delta$ to finite orders and setting $\delta=1$ in the end of the calculation.  Thus, our approach has some similarities with the ordinary delta expansion, and we use the term "delta expansion" to refer our method.   We emphasize that our delta expansion on the lattice needs no extra parameter and the principle of minimum sensitivity \cite{ste}, both of which play, in the ordinary delta expansion, important but somewhat artificial roles to produce non-trivial results. 

To investigate and explore the above idea, we apply our method to the $d=2$ non-linear $\sigma$ model on the lattice, which model is also called $N$-vector model.  In the present paper, we focus on two extreme cases,  $N=\infty$ and $N=1$ which case corresponds to the Ising model.   The model can be exactly solved in the large $N$ limit, so we can examine to what extent our proposal is effective both in the qualitative and quantitative sense.   The $N=1$ case is also of interest, since the Ising model at $d=2$ serves us a good testing ground of analyzing phase transition at non-zero temperature.

\section{Delta expansion in simple examples}\quad 
To illustrate our strategy, we first study two simple examples.
\subsection{Example 1}  Consider the problem of approximating the value of $\Omega(0)$ where $\Omega(x)$ is given as a finite series in $1/x$;
$$
\Omega(x)=\frac{1}{x}-\frac{1}{x^2}+\frac{1}{x^3}-\cdots\left(=\frac{1}{1+x}\right).
$$

We introduce $\delta$ by the replacement of $x$ to $x(1-\delta)$ ($0\le \delta \le 1$) in $\Omega(x)$, resulting a new function of two variables $\delta$ and $x$, $\Omega(x(1-\delta))$.   Note that we can make the region around $x=0$ of $\Omega(x)$ wider by setting the value of $\delta$ as close to $1$.   Then, if we can construct a truncated series of $\Omega(x(1-\delta))$ in $1/x$ and $\delta$ and it is effective at $x>x^{*}\sim O(1)$, the series would provide us the information on $\Omega(0)$ as long as $\delta\sim 1$.  Here, we stress that for the plan to work the factor $1-\delta$ must be expanded in $\delta$.   As a systematic expansion of $\Omega(x(1-\delta))$ effective at large $x$ and small $\delta$, we employ the ordinary expansion in both variables around $1/x=\delta=0$.   Then $n^{\rm th}$ order approximant of $\Omega(x(1-\delta))$, denoted as $\tilde \Omega_{n}(x,\delta)$, is written formally by
$$
\tilde \Omega_{n}(x,\delta)=c+\sum_{\mu}c_{\mu}z_{\mu}+\cdots+\sum_{\mu_{1},\mu_{2},\cdots,\mu_{n}}c_{\mu_{1},\mu_{2},\cdots,\mu_{n}}z_{\mu_{1}}z_{\mu_{2}}\cdots z_{\mu_{n}},
$$
where $z_{1}=1/x$ and $z_{2}=\delta$.  For example, when $n=3$, we have
$$
\tilde\Omega_{3}(x,\delta)=\frac{1}{x}(1+\delta+\delta^2)-\frac{1}{x^2}(1+2\delta)+\frac{1}{x^3}.
$$
In general, it is efficient to obtain the above expansion in the following manner:  Consider
$$
\Omega_{n}(x)=\frac{1}{x}-\frac{1}{x^2}+\cdots+\frac{(-1)^{n-1}}{x^{n}}.
$$
First shift $x$ to $x(1-\delta)$ and expand $\Omega_{n}(x(1-\delta))$ in $\delta$ to the relevant order.  For example the term $x^{-r}$ should be expanded to the order $\delta^{n-r}$, giving $x^{-r}\to x^{-r}(1+r\delta+\cdots+\frac{(n-1)!}{(r-1)!(n-r)!}\delta^{n-r})$.  Setting $\delta=1$ which means the infinite dilation, we then have 
\begin{equation}
\frac{1}{x^{r}}\to \frac{n!}{r!(n-r)!}\frac{1}{x^r}=\bigg(
\begin{array}{c}
n \\
r
\end{array}
\bigg)\frac{1}{x^r}=D\Big[\frac{1}{x^r}\Big].
\label{1}
\end{equation}
From now on we use the symbol $D$ as the operation of delta expansion to the relevant order with setting $\delta=1$ understood.  
To the order $n$ we therefore obtain
\begin{eqnarray*}
\tilde\Omega_{n}(x,1)&=&D\left[\frac{1}{x}-\frac{1}{x^2}+\frac{1}{x^3}-\cdots+\frac{(-1)^{n-1}}{x^{n}}\right]
=\sum_{r=1}^{n}(-1)^{r-1}\bigg(
\begin{array}{c}
n \\
r
\end{array}
\bigg)\frac{1}{x^r}\\
&=&1-\Big(1-\frac{1}{x}\Big)^{n}.
\end{eqnarray*}
It is easy to see that $\left(1-\frac{1}{x}\right)^{n}\to 0 \,\,(n\to \infty)$ for $x>1/2$.  
Thus, when $x>1/2$,
$$
\tilde\Omega_{n}(x,1)\to 1=\Omega(0).
$$
FIG. 1 shows the function $\tilde\Omega_{n}(x,1)$ at $n=2,3,\cdots,10$.    At those orders, there exists a plateu and the region of the plateu grows wider as the order increases.  The value at a flat point which is a typical value on the plateu agrees with the value,  $\Omega(0)=1$.   Thus the result is quite satisfactory.
\begin{figure}[htb]
\begin{center}
\includegraphics[scale=0.65]{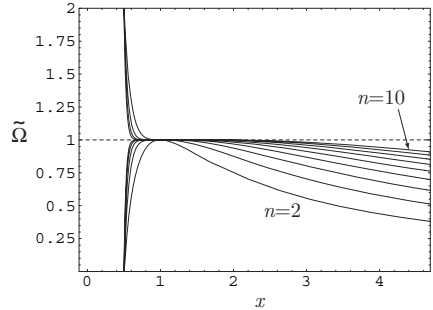}
\caption{Plots of $\tilde\Omega_{n}(x,1)$ at $n=2,3,\cdots,10$.  The dotted line represents the value $\Omega(0)=1$.}
\end{center}
\end{figure}

\subsection{Example 2}Next example deals with the relation of observables appearing in the $d=1$ Ising model.  In the model, the inverse temperature $\beta$ is related to $M$, the square of the screening mass  (in lattice units)  in the momentum representation, by 
$$
\beta=\frac{1}{4}\log\left(1+\frac{4}{M}\right).
$$
At very low temperature the mass $M$ is very small and we have the logarithmic relation
\begin{equation}
\beta\sim \frac{1}{4}\log\frac{4}{M},
\label{2}
\end{equation}
while at high temperature, $M\gg 1$ and 
\begin{equation}
\beta=\frac{1}{M}-\frac{2}{M^2}+\frac{16}{3M^3}-\frac{16}{M^4}+\cdots.
\label{3}
\end{equation}
Though the previous example has the limit, $\lim_{x\to 0}\Omega(x)=1$, 
in the present example, $\beta$ diverges logarithmically in the $M\to 0$ limit.  We show, however, that the delta expansion on the "high temperature expansion" (\ref{3}) recovers numerically the asymptotic behavior of $\beta$ near $M\sim 0$ represented by (\ref{2}).

\begin{figure}[htb]
\begin{center}
\includegraphics[scale=0.65]{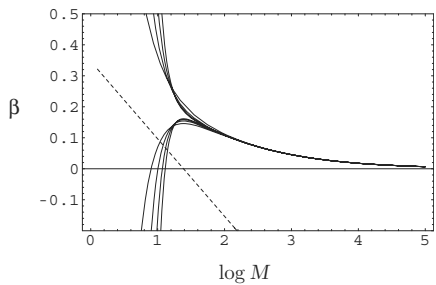}
\caption{Behaviors of the "high temperature series" of $\beta$ from $3^{\rm rd}$ to $10^{\rm th}$ orders.  The dotted line represents the leading small $M$ behavior.}
\end{center}
\end{figure}
The series (\ref{3}) itself breaks down at $M=4$ (see FIG. 2) and it cannot be used for studying the small $M$ behavior of $\beta$.  To improve the status, we implement the dilation of small $M$ region around $M=0$ by shifting $M\to  (1-\delta)M$ in $\beta(M)$.   Using (\ref{1}), we then perform delta expansion at large $M$ to give
$$
\tilde\beta_{n}=\bigg(
\begin{array}{c}
n \\
1
\end{array}
\bigg)\frac{1}{M}-\bigg(
\begin{array}{c}
n \\
2
\end{array}
\bigg)\frac{2}{M^2}+\bigg(
\begin{array}{c}
n \\
3
\end{array}
\bigg)\frac{16}{3M^3}-\bigg(
\begin{array}{c}
n \\
4
\end{array}
\bigg)\frac{16}{M^4}+\cdots.
$$  
To examine whether the above series captures the scaling behavior, we study the modification of the small $M$ behavior due to the delta expansion.

At small $M$, $\beta$ is expanded as $\beta=\frac{1}{4}\log\frac{4}{M}+\frac{1}{4}(\frac{M}{4}-\frac{M^2}{32}+\cdots)$.  The leading term changes as
$$
\frac{1}{4}\log\frac{4}{(1-\delta)M}=\frac{1}{4}\log\frac{4}{M}+\frac{1}{4}(\delta+\frac{1}{2}\delta^2+\frac{1}{3}\delta^3+\cdots).
$$
Truncating at $\delta^n$ and setting $\delta=1$, we have
$$
\frac{1}{4}\log\frac{4}{(1-\delta)M}\sim \frac{1}{4}\log\frac{4}{M}+\frac{1}{4}\sum_{k=1}^{n}\frac{1}{k}=D\bigg[\frac{1}{4}\log\frac{4}{M}\bigg].
$$
Note that as $n\to \infty$, the constant part $\sum_{k=1}^{n}\frac{1}{k}$ diverges as $\log n$.  This reflects the logarithmic divergence of $\beta(M)$ in the $M\to 0$ limit.   The corrections by terms of positive integer powers experience drastic change:  When the delta expansion is performed to a large order, lower order terms vanish when the value of $\delta$ is tuned to $1$.  For example, we find that $m\to m(1-\delta)$, $m^2\to m^2(1-2\delta+\delta^2)$ and $m^3\to m^3(1-3\delta+3\delta^2-\delta^3)$, and these terms vanish themselves at orders large enough.   At small $M$, however,  the delta expansion has a subtlety on the definition of the full order that is in accordance with the definition at large $M$.   Though we do not know a convincing definition, let us proceed by supposing that  we should confine ourselves with only the leading term at small $M$ and it should be expanded in $\delta$ to the same order with the full order at large $M$. 
  Thus, at $n^{\rm th}$ order of delta expansion, $\tilde\beta_{n}$ behaves at small $M$ as
\begin{equation}
\tilde \beta_{n}\sim \frac{1}{4}\left(\log\frac{4}{M}+\sum_{k=1}^{n}\frac{1}{k}\right).
\label{3a}
\end{equation}

As is obvious from the plot of $\tilde\beta_{n}$ (see FIG.3), the logarithmic behavior (\ref{3a}) is recovered by the $\delta$- expanded $1/M$ series already at $n=3$.   Note that the scaling region develops to larger $M$ as the order of expansion grows.  In the scaling region to be seen in FIG.3, small $M$ behavior of $\beta(M)$ represented by (\ref{2}) is well approximated by subtracting $\frac{1}{4}\sum\frac{1}{k}$ from $\tilde{\beta}$.  Thus, even the limit of sequence $\{\tilde\beta_n\}$ does not exist for any $M$ as $n\to \infty$, we can reproduce the scaling behavior both in the qualitative and quantitative respects.
\begin{figure}[htb]
\begin{center}
\includegraphics[scale=0.65]{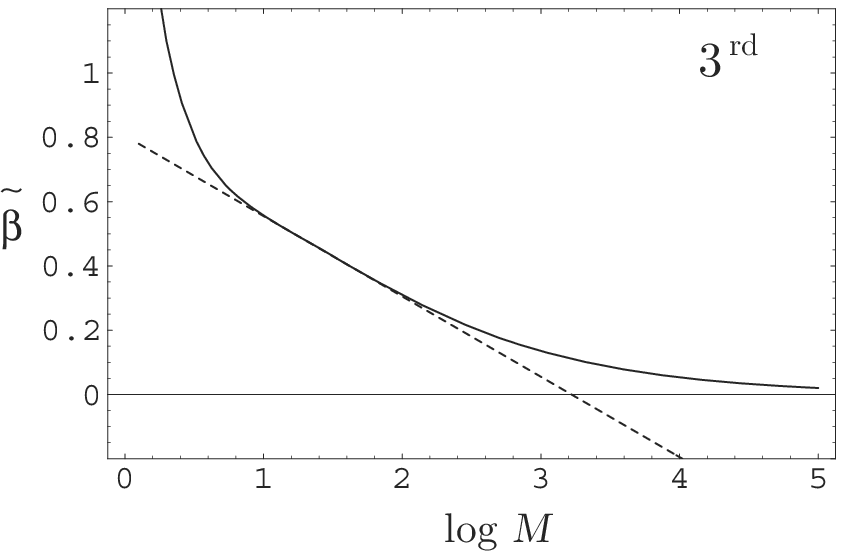}
\includegraphics[scale=0.65]{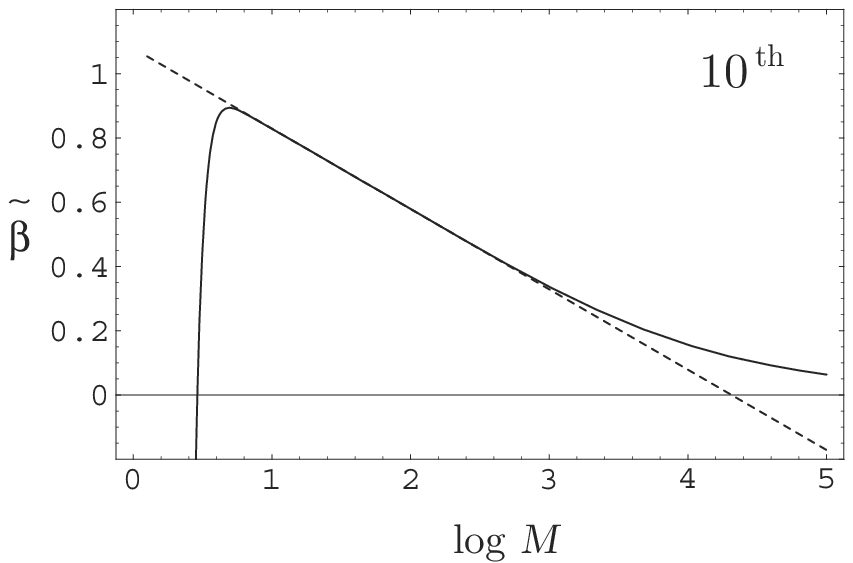}
\caption{Behaviors of the $\delta$-expanded $1/M$ series of $\tilde\beta$ at $3^{\rm rd}$ 
and $10^{\rm th}$ orders.  The dotted lines represent the leading small $M$ behaviors at respective orders.}
\end{center}
\end{figure}

\section{The non-linear $\sigma$ model at large $N$}
\subsection{Brief review of the model}
The non-linear $\s$ model at
two-dimensional Euclidean space is defined by the action,
$$
{\cal L}={1 \over 2f}\sum_{\mu}\big(\partial_{\mu} \vec\s\big)^2,
$$
where the fields $\s^{i}(x)\,\,(i=1,2,\cdots,N)$ obey the constraint,
$$
\vec{\s}^2(x)=\sum_{i=1}^{N}\s^{i}(x)\s^{i}(x)=N.
$$
The discretized space we work with is the periodic square lattice of the lattice spacing
$a$ where sites are numbered by two integers, $(n_{1},n_{2})={\bf n}$.  On the lattice the action may be written as
\begin{equation}
S=2\beta \sum_{\bf n}\vec{\s}_{\bf n}^2-\beta\sum_{\bf n}\sum_{\mu=1,2}\vec{\s}_{\bf n}\cdot\vec{\s}_{\bf n+\bf e_{\mu}},
\label{4}
\end{equation}
where $\beta$ is defined as the inverse of the bare coupling constant $f$,
$$
\beta:=\frac{1}{f}
$$
and 
$\vec{\s}_{\bf n+\bf e_{\mu}}$ stands for the nearest neighbour spin of $\vec{\s}_{\bf n}$ with ${\bf e}_{1}=(1,0)$ and ${\bf e}_{2}=(0,1)$.  The constraint is same as that in the continuum case,
$\vec{\s}_{\bf n}^2=N$, and the first term in (\ref{4}) is actually a constant that can be omitted. 

Consider the correlation of fields, 
$$\beta \la \vec{\s}_{\bf{0}}\cdot \vec{\s}_{\bf{r}}\ra=N\int^{\pi/a}_{-\pi/a}\frac{d^{2}{\bf p}}{(2\pi)^2}\exp(-i{\bf p}\cdot{\bf r}a)G({\bf p},\beta, a).
$$ 
By the use of Fourier transform method, one can calculate $G({\bf p}, \beta, a)$ to higher orders in $\beta$.   In particular, two moments $\sum_{\bf n}\la\s_{\bf 0}\s_{\bf n}\ra$ and $\sum_{\bf n}{\bf n}^2\la  \s_{\bf 0}\s_{\bf n}\ra$ are calculated to $21^{\rm th}$ order \cite{but}.  One can use the result to obtain the correlation length $\xi$ as a series in $\beta$.   
In the large $N$ limit, however, it is more convenient for us to utilize the constraint and the fact that the structure of $G$ becomes simple due to the absence of the wave function renormalization.   Namely, taking the constraint $\vec{\s}^2=N$ into account and setting ${\bf r}={\bf 0}$ in the correlation $\la \vec{\s}_{\bf 0}\cdot\vec{\s}_{\bf r}\ra$, we have 
\begin{equation}
\beta=\int^{\pi}_{-\pi}\frac{d^{2}{\bf p}}{(2\pi)^2}\frac{1}{M+2\sum_{\mu=1,2}(1-\cos p_{\mu})}.
\label{5}
\end{equation} 
Here we have rescaled the momentum by ${\bf p}\to{\bf p}/a$ and defined $M$ by
$$
M=m^2 a^2,
$$
where $m$ represents the physical mass.  Note that $M$ is related to $\xi$ by $M=\xi^{-2}$.  Since we consider in the next subsection the dilation of the region of $M$ around $M=0$, we express the inverse coupling $\beta$ as a function of $M$.

When the lattice spacing is small enough where $M\ll 1$, we find from (\ref{5})
\begin{equation}
\beta=-\frac{1}{4\pi}\log\frac{M}{32}+\frac{M}{32\pi}\Big(\log\frac{M}{32}+1\Big)+O(M^2\log M).
\label{6}
\end{equation} 
Keeping only the first term, we have
$$
m^2=32 a^{-2}\exp(-\frac{4\pi}{f})=32 \Lambda_{L}^2.
$$
This represents the dynamical mass in terms of the mass scale $\Lambda_{L}$ defined on the lattice.   
On the otherhand, when the lattice spacing is large where $M\gg1 $, straightforward expansion of the right hand side of (\ref{5}) in $M^{-1}$ gives the following:
\begin{equation}
\beta={1 \over M}-{4 \over M^2}+{20 \over
M^3}-{112
\over M^4}+{676 \over M^5}-{4304
\over M^6}+{28496 \over
M^7}-O(M^{-8}).
\label{7}
\end{equation}
\begin{figure}[h]
\begin{center}
\includegraphics[scale=0.65]{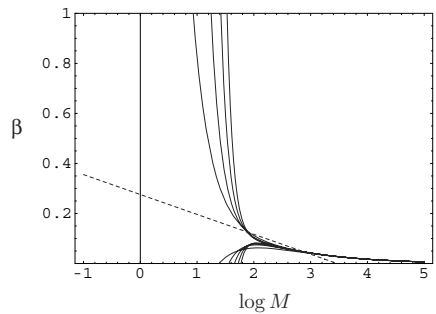}
\caption{The graphs of $\beta(=1/f)$ as the function of $\log M$.  The dashed line represents
the asymptotic behavior at small enough $M$,  $\beta=-\frac{1}{4\pi}\log\frac{M}{32}$.  Solid curves
represent the graphs in the large $M$ expansion from $2^{\rm nd}$ to
$9^{\rm th}$ orders. }
\end{center}
\end{figure}
From FIG. 4, it
is apparent that the series (\ref{7}) breaks down around $\log M\sim 2$ and the 
scaling behavior, $\beta\sim -\frac{1}{4\pi}\log\frac{M}{32}$, is not observed as it would. 

\subsection{Delta expansion}
We perform the delta expansion of $\beta(M)$ with respect to $a^2$ or $M$ following the manner adopted in example 2 studied in section II.  We shall see the asymptotically free behavior in the large $M$ expansion of $\tilde\beta$ and estimate the value of the non-perturbative dynamical mass $m$ in units of the lattice lambda parameter.

 From (\ref{1}), we find
\begin{equation}
\tilde\beta_{n} =\bigg(
\begin{array}{c}
n \\
1
\end{array}
\bigg)\frac{1}{M}-\bigg(
\begin{array}{c}
n \\
2
\end{array}
\bigg)\frac{4}{M^2}+\bigg(
\begin{array}{c}
n \\
3
\end{array}
\bigg)\frac{20}{M^3}-\cdots+\bigg(
\begin{array}{c}
n \\
n
\end{array}
\bigg)\frac{const}{M^n}.
\label{8}
\end{equation}
At small $M$, the leading term is transformed to
$$
\tilde\beta_{n}\sim D\bigg[-\frac{1}{4\pi}\log\frac{M}{32}\bigg]=-\frac{1}{4\pi}\Big(\log\frac{M}{32}-\sum_{k=1}^{n}\frac{1}{k}\Big).
$$

We examine whether the scaling behavior written above is seen in (\ref{8}) or not.   FIG.5 shows the plots of $\tilde\beta_{n}$ at $n=4,9, 15 $ and $20$.  Dashed lines represent the leading small $M$ behaviors at respective orders.  
\begin{figure}[h]
\begin{center}
\includegraphics[scale=0.65]{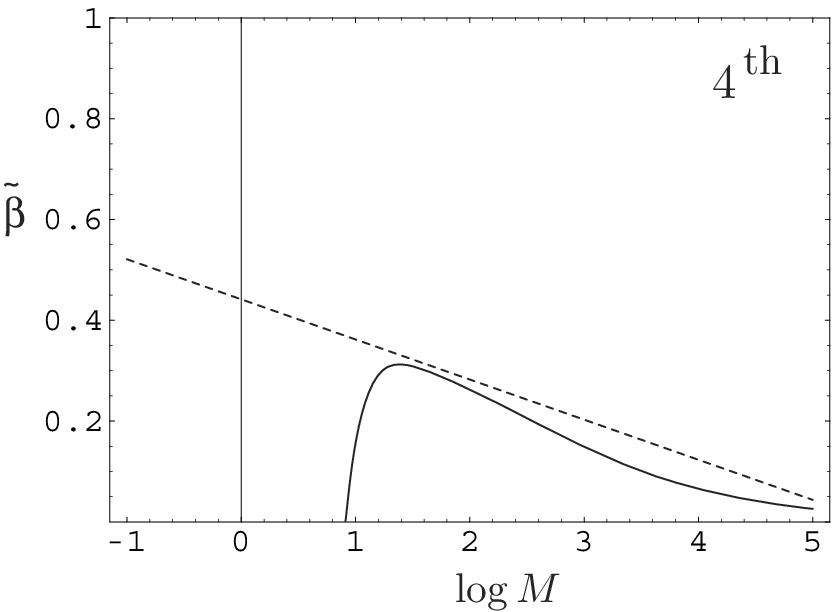}
\includegraphics[scale=0.65]{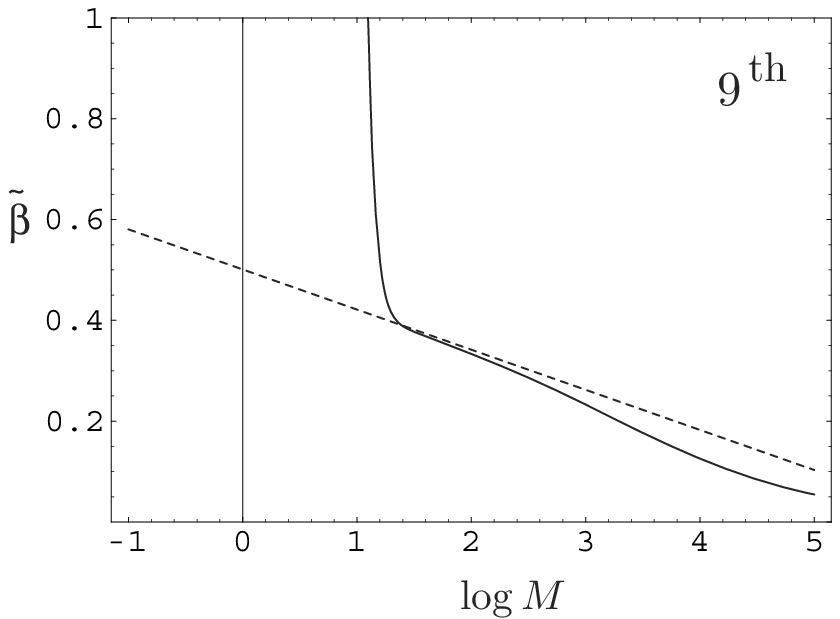}
\includegraphics[scale=0.65]{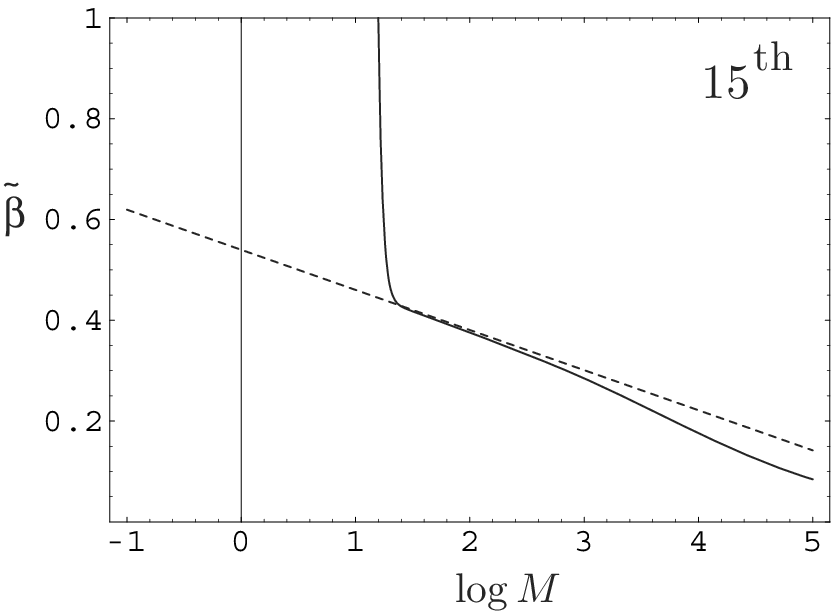}
\includegraphics[scale=0.65]{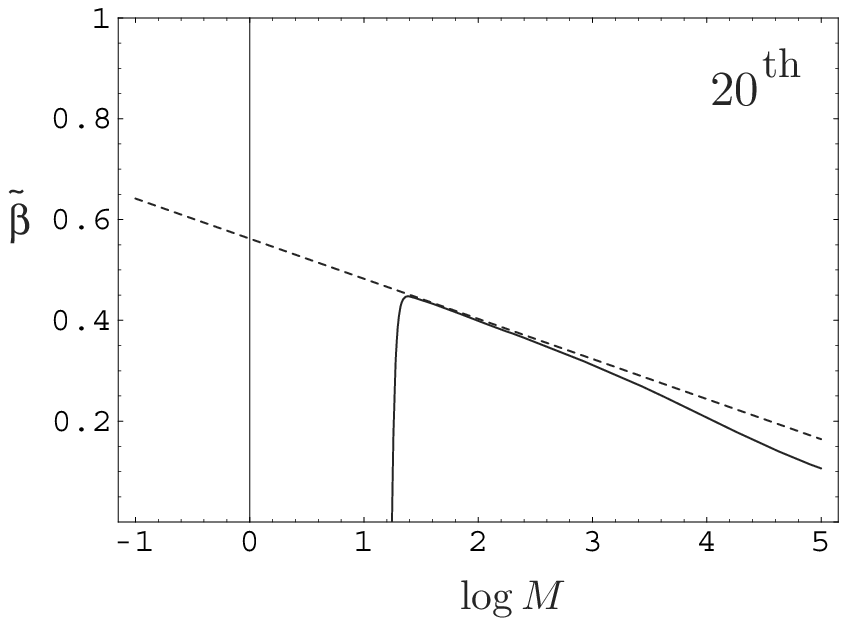}
\caption{Graphs of $\tilde\beta_{n}$ at $4^{\rm th}$, $9^{\rm th}$,  $15^{\rm th}$ and $20^{\rm th}$ orders as the function of $\log M$.  The dashed lines represent the leading asymptotic behaviors at small $M$ at respective orders.}
\end{center}
\end{figure}
We see that $\tilde\beta_{n}(M)$ in $1/M$ expansion is effective to $\log M\sim 1$.  In addition, we find that the logarithmic scaling behavior, $\tilde\beta_{n}\sim -\frac{1}{4\pi}\log M$, is seen at several and higher orders around $\log M\sim 2$.  Due to the dilation, the scaling region realized in $1/M$ expansion (\ref{8}) may develop toward larger $M$ region.  The tendency is confirmed from FIG. 5.

Having observed the asymptotic scaling behavior in $1/M$ expansion, we can estimate the mass $m$ in terms of $\Lambda_{L}$ even when the knowledge on the detailes of the small $M$ behavior such as the information of the constant $\log 32$ is absent.  Let us write the scaling behavior as
\begin{equation}
\beta\sim-\frac{1}{4\pi}(\log M+C).
\label{9}
\end{equation}
Note that (\ref{9}) is derived only from the ultraviolet structure of the model.  The value of the constant $C (=-\log 32)$ is, however, not obtained by the renormalization group argument alone.  Then we like to show that our approach enables one to obtain the approximate value of the constant.

We look for the matching point where the asymptotic behavior is supposed to  begin and from there extrapolate $\tilde\beta(M)$ (at large $M$) to small $M$ region.  The matching point may be fixed by requiring the agreement of the derivative of $\tilde\beta$ at large $M$ with that of the leading term at small $M$ which comes from renormalization group argument;
$$
\frac{\partial \,\tilde\beta(M)|_{large\, M}}{\partial \log M}=\frac{\partial \,\tilde\beta(M)|_{scaling}}{\partial \log M}=-\frac{1}{4\pi}.
$$
Up to $20^{\rm th}$ orders, the solutions, $M=4.888,\,4.7539,\,4.6479,\,\cdots, \, 4.3597$ exist at $n=4,6,8,\cdots, 20$ respectively.   The value of $C$ can be estimated by
$$
-\frac{C}{4\pi}=\bigg[\tilde\beta_{n}(M)|_{large\,M}+\frac{1}{4\pi}(\log M-\sum_{k=1}^{n}\frac{1}{k})\bigg]_{M=M^{*}},
$$
where $M^{*}$ denotes the solution.  The result is $-3.3066, -3.3643, -3.3918, \cdots, -3.4383$ at $n=4,6,8,\cdots,20$.  Then, 
 the extrapolated asymptotic behavior, for example at $n=4$, predicts
\begin{equation}
\beta \big |_{scaling}\sim  -\frac{1}{4\pi}(\log M-\log 27.29).
\label{10}
\end{equation}
Comparison of (\ref{10}) to (\ref{9}) with $C=-\log 32$ tells us that we have $\log\frac{M}{27.29}$ for $\log \frac{M}{32}$.  Now the approximation of the logarithmic constant leads us to the approximation of the dynamical mass.  The result is $m^2\approx 27.29\Lambda_{L}^2$ at $4^{\rm th}$ order.   In the same manner, we obtain the ratio $m^2/\Lambda_{L}^2=28.91,\,29.72,\,30.20,\cdots,\,31.13$ at $n=6,8,10,\cdots, 20$.  Better values are obtained at higher orders and the sequence implies the convergence to the exact value.

\subsection{Symanzik improvement}
In this subsection, we consider how to accerelate the speed of approaching to the asymptotic scaling and raise the accuracy of estimating  the dynamical mass.  

The clue to the resolution is to note the presence of the subleading logarithmic terms, $M^{k}\log M$, in $\beta(M)$ at small $M$ (see (\ref{6})).  
The second and higher order contributions in (\ref{6}) may prevent from the dominance of the leading term as we can see below:   
Consider the effect of these subleading logs in the application of the delta expansion ton $\beta(M)$ at small $M$.   If we expand powers and logarithms of $M$ to $\delta^{n}$, we find
$$
D [1]=1,\qquad D[ M^{r}]=0\quad (1\le r\le n)
$$
and
$$
D[\log M]=\log M-\sum_{k=1}^{n}\frac{1}{k},\quad D [M\log M]=-\frac{M}{n},\quad D[M^2\log M]=2!\frac{M^2}{n^2},\quad \cdots.
$$
It is important to note that, though the terms of positive integer powers vanish, the logarithmic corrections survive after the delta expansion.  Then suppose that $M$ is large enough so that we can neglect  the problem that to which order the first few terms should be expanded in $\delta$.  We then have
$$
\tilde\beta(M)\big|_{small\, M}\sim-\frac{1}{4\pi}\Big(\log\frac{M}{32}-\sum_{k=1}^{n}\frac{1}{k}-\frac{M}{8n}+O(M^2)\Big).
$$
The third term of order $M$ delays the scaling of $\tilde\beta$ at finite $n$.

The origin of the subleading logarithmic corrections is the propagator modified on the lattice.  Actually, 
the expansion of the propagator at small $p^2$ reads
\begin{equation}
\frac{1}{M+\sum_{\mu}p_{\mu}^2-\frac{1}{12}\sum_{\mu}p^{4}+\cdots}=\frac{1}{M+\sum_{\mu}p_{\mu}^2}+\frac{1}{12}\frac{\sum_{\mu}p^{4}}{(M+\sum_{\mu}p_{\mu}^2)^2}+\cdots 
\label{11}
\end{equation}
and the momentum integration yields $-\frac{1}{4\pi}\log M$ from the first term and $\frac{1}{32\pi}M\log M$ from the second term.  
In general, $k^{\rm th}$ term gives $const. M^{k-1}\log M+{\rm regular \hskip 4pt terms \hskip 4pt in}\hskip 4pt M$.  Thus, it is highly expected that Symanzik improvement \cite{sym} accerelates the quick dominance since it subtracts the higher order $p^2$ corrections in the propagator. 

At the first order of Symanzik imporovement scheme, the coupling of spins at next-to-the nearest neighbour sites in both directions are introduced.  It is well known that the resulting action becomes
$$
S=\beta\sum_{\bf n}
\bigg[\frac{5}{2}\vec{\s}_{\bf n}^2-\frac{4}{3}\sum_{\mu}\vec{\s}_{\bf n}\cdot\vec{\s}_{{\bf n}+{\bf e}_\mu}+\frac{1}{12}
\sum_{\mu}\vec{\s}_{\bf n}\cdot\vec{\s}_{{\bf n}+2{\bf e}_\mu}\bigg].
$$
Also in the improved action, we have the following result of the constraint at large $N$,
\begin{equation}
\beta=\int_{-\pi}^{\pi}\frac{d^2 {\bf p}}{(2\pi)^2} \frac{1}{
M+5-\frac{8}{3}\sum_{\mu}\cos p_{\mu}+\frac{1}{6}\sum_{\mu}\cos 2p_{\mu} }.
\label{12}
\end{equation}
Expansion of the denominator of the propagator for small $p_{\mu}$ gives
$$
M+\sum_{\mu}p_{\mu}^2+O(p^6)
$$
which has no $p^4$ contribution and thus $M\log M$ term is absent in $\beta$ at small $M$.   From (\ref{12}), we find that $\beta$ behaves at small $M$ as
$$
\beta=-\frac{1}{4\pi}\big(\log M+C'\big)+O(M),
$$
where $C'=-2.87298$.   This should be compared with (\ref{9}).  Though $\log M$ is invariant, the constant part which is not of universal nature is modified from $C$ to $C'=C+0.592754$.  The next-to-the leading logarithmic term, $M\log M$, is absent and the approach to the asymptotic scaling would become faster than before because the correction to the scaling of $\tilde\beta$ is at most $O(M^2)$.

From (\ref{12}) we find the large $M$ expansion of $\beta$,
\begin{equation}
\beta= \frac{1}{M}- 
  \frac{5}{M^2}+ \frac{1157}{36 M^3}  - 
  \frac{8419}{36 M^4}+ \frac{3190499}{1728 M^5}+O(M^{-6}).
  \label{13}
\end{equation}
\begin{figure}[h]
\begin{center}
\includegraphics[scale=0.65]{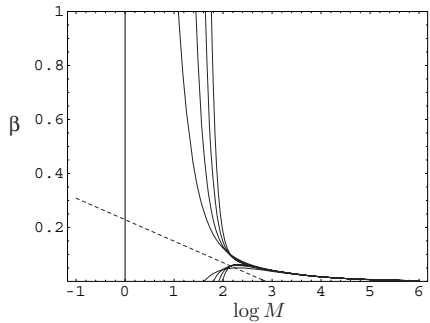}
\caption{The graphs of the large $M$ series of $\beta$ in the first order Symanzik's improved action from $2^{\rm nd}$ to $9^{\rm th}$ orders}
\end{center}
\end{figure}
As is clear from FIG. 6, the above series is valid only for large $M$ as in the case of the previous series (\ref{7}).  However, once delta expansion is applied, we find that the large $M$ series exhibits the correct logarithmic scaling behavior also for the improved action (see FIG. 7).  Moreover, the extrapolated scaling function produces the following good approximate values for $C'=-2.87298$,
$$
-2.8359,\quad -2.8555,\quad -2.8629,\quad \cdots,\quad -2.8713,
$$
for orders $n=4,\,6,\,8,\cdots,\,20$, respectively.   The accuracy is much improved from those of the original action.
\begin{figure}[h]
\begin{center}
\includegraphics[scale=0.65]{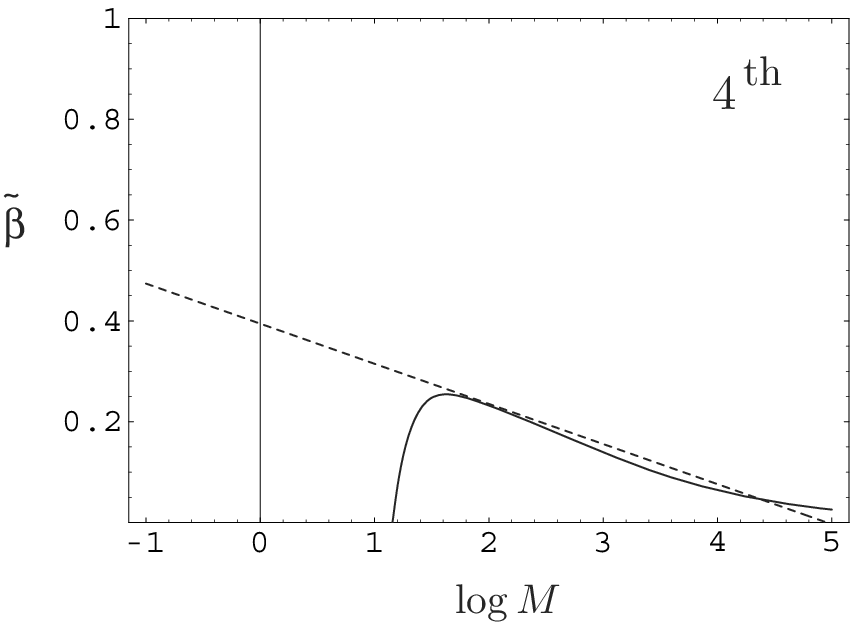}
\includegraphics[scale=0.65]{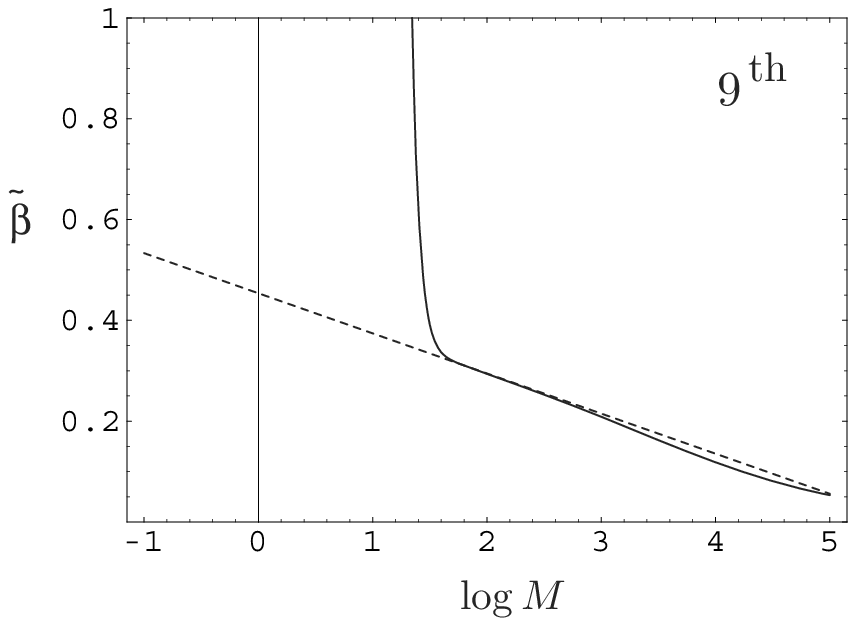}
\includegraphics[scale=0.65]{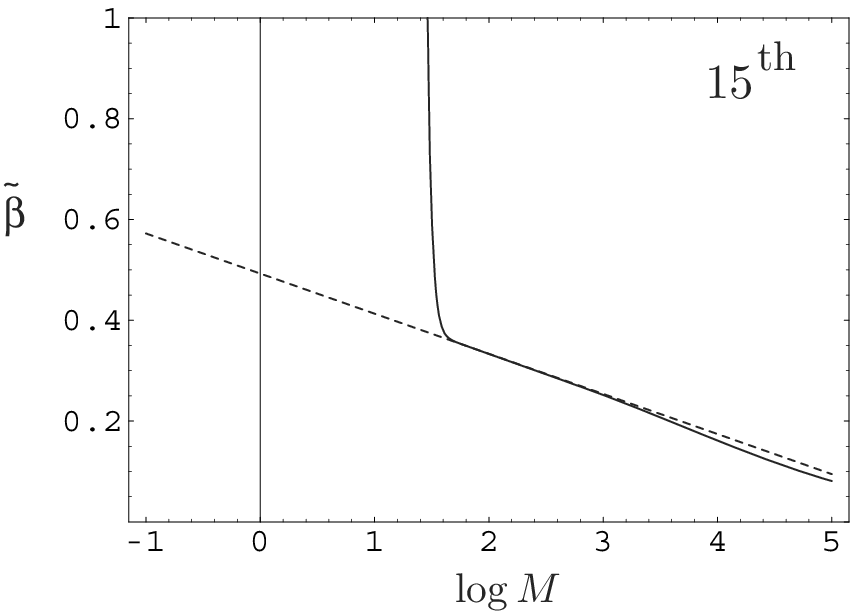}
\includegraphics[scale=0.65]{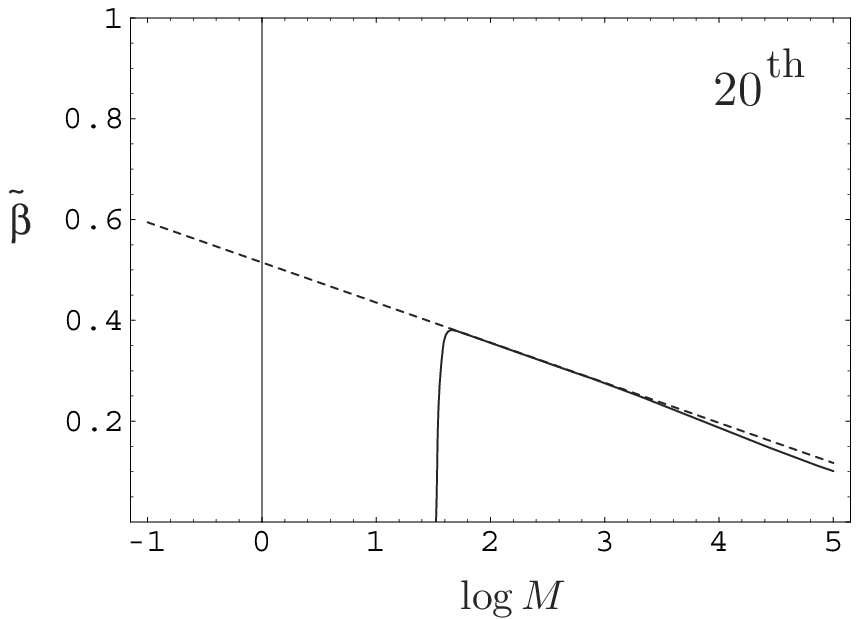}
\caption{Graphs of $\tilde\beta_{n}$ at $4^{\rm th}$, $9^{\rm th}$,  $15^{\rm th}$ and $20^{\rm th}$ orders as a function of $\log M$ in the first order Symanzik action.  The dashed lines represent the leading asymptotic behaviors at small $M$ at respective orders.}
\end{center}
\end{figure}

To study further the results of Symanzik's improvement, we proceed to the second order.   By
introducing the spin-spin coupling of the form $\vec{\s}_{\bf n}\cdot\vec{\s}_{{\bf n}+3{\bf e}_\mu}$ with the suitable weight, we can eliminate the second logarithmic correction $M^2\log M$ in
small $M$ expansion of $\beta$.   The constraint equation becomes 
$$
\beta=\int_{-\pi}^{\pi}\frac{d^2 {\bf p}}{(2\pi)^2} \frac{1}{
M+\frac{49}{9}-\frac{4}{3}\sum_{\mu}\cos p_{\mu}+\frac{3}{10}\sum_{\mu}\cos 2p_{\mu} -\frac{1}{45}\sum_{\mu}\cos 3p_{\mu}}
$$
and the scaling behavior of $\beta$ is obtained as
$$
\beta=-\frac{1}{4\pi}(\log M+C'')+O(M),
$$
where $C''=-2.72121=C+0.744526$. 

Now the large $M$  expansion of $\beta(M)$ gives
$$
\beta=\frac{1}{M}-\frac{49}{9M^2}+\frac{313733}{8100 M^3}-\frac{2288857}{7290 M^4}+\frac{727664156617}{262440000 M^5}-O(M^{-6}).
$$
As in the case of the first order improvement, the scaling behavior is observed in the $\delta$-expanded large $M$ series and the extrapolation gives the approximation of the constant $C''=-2.72121$.   At $n=4,\,6,\,8,\,\cdots,\,20$, the results are as follows:
$$
-2.70458,\quad -2.71498,\quad -2.71823,\quad -2.71956,\quad \cdots,\quad -2.72097.
$$
The accuracy was further improved.  We conclude that Symanzik's action plays a crucial role in the quantitative improvement on the delta expansion approach.

\section{Ising model at $d=2$}
Next we turn to discuss the Ising model at $d=2$ which corresponds to $N=1$ case of the $N$-vector model.   As well known, the second order phase transition at non-zero temperature is driven by the correlation length grown to infinitely large.  Therefore the transition may be analyzed by dilation and delta expansion around the massless limit.   

From the calculation of the correlation function, the correlation length $\xi$ or $M=\xi^{-2}$ would be obtained as a function of $\beta$, the inverse temperature.  In the Ising case, $\sum_{\bf n}\la\s_{\bf 0}\s_{\bf n}\ra$ and $\sum_{\bf n}{\bf n}^2\la  \s_{\bf 0}\s_{\bf n}\ra$ were calculated  in \cite{but2} to $25^{\rm th}$ order in $\beta$.  We use the result and find that 
$$
M=\frac{1}{2\beta}-2+\frac{5\beta}{3}+\frac{67\beta^3}{45}+\frac{38\beta^5}{189}+\frac{9697\beta^7}{4725}-16\beta^8+\frac{1035164\beta^9}{18711}-\frac{352\beta^{10}}{3}+\cdots.
$$
By inverting $M^{-1}$ and $\beta$, we obtain
\begin{equation}
\beta=\frac{1}{2M}-\frac{1}{M^2}+\frac{29}{12M^3}-\frac{13}{2M^4}+\frac{1503}{80M^5}-\frac{1373}{24M^6}+\frac{40581}{224M^7}-\frac{9461}{16M^8}+\frac{4553741}{2304M^9}-\frac{4309333}{640M^{10}}+\cdots.
\end{equation}

Now, near the transition point, conventional scaling form reads that
$$
\xi\sim const.\left(\frac{\beta_{c}}{\beta}-1\right)^{-\nu}
$$
where the critical exponent $\nu$ and the inverse of the critical temperature $\beta_{c}$ are known to have values, $\nu=1$ and $\beta_{c}=\frac{1}{2}\log(1+\sqrt{2})$ \cite{itz}.   Using $\xi^2=M^{-1}$, we then have
\begin{equation}
\beta\sim \beta_{c}-A M^{\frac{1}{2\nu}}.
\label{14}
\end{equation}
From (\ref{14}), we see that the present case has the limit $\lim_{M\to 0}\beta(M)=\beta_{c}$.  However, the derivative of the first correction diverges as $M\to 0$ since the power of $M$, $\frac{1}{2\nu}$ is smaller than $1$.  This makes the convergence of $\beta(M)$ to $\beta_{c}$ slow and forces us to have $1/M$ series of $\beta$ to very large orders for the precise evaluation of $\beta_{c}$ (see FIG. 8).   
\begin{figure}[h]
\begin{center}
\includegraphics[scale=0.65]{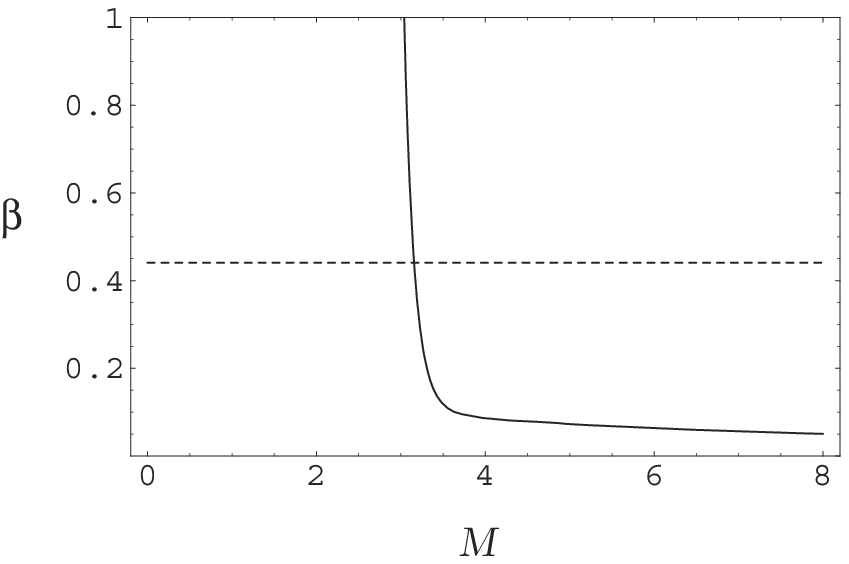}
\includegraphics[scale=0.65]{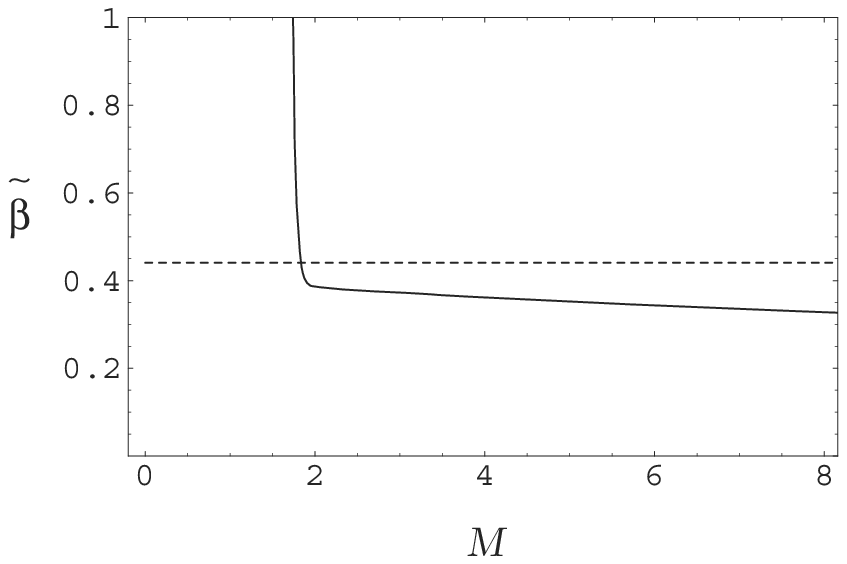}
\caption{$\beta$ and $\tilde\beta$ as functions of $M$ at $25^{\rm th}$ order.  The dotted lines represent $\beta_{c}=\frac{1}{2}\log(1+\sqrt{2})$.}
\end{center}
\end{figure}  
Then, rather than $\beta_{c}$, we turn to the estimation of the critical exponent $\nu$ .  For the purpose, we consider $\frac{\partial}{\partial\log M}\log[-\frac{\partial\beta }{\partial\log M}]$ of which behavior at scaling region reads
\begin{equation}
\phi(M)=\frac{\partial}{\partial\log M}\log[-\frac{\partial\beta }{\partial\log M}]\sim \frac{1}{2\nu}.
\label{15}
\end{equation}
The leading term of $\phi(M)$ explicitly written in the right hand side of (\ref{15}) comes from the second term of (\ref{14}).  Since the leading term of $\phi$ is independent of $M$ and invariant under the delta expansion, we expect that $\tilde\phi=\frac{\partial}{\partial\log M}\log[-\frac{\partial\tilde\beta }{\partial\log M}]$ also behaves at small $M$ as $\tilde\phi\sim \frac{1}{2\nu}$.   Here $\tilde\beta$ is given by
\begin{equation}
\tilde\beta_{n} =\bigg(
\begin{array}{c}
n \\
1
\end{array}
\bigg)\frac{1}{2M}-\bigg(
\begin{array}{c}
n \\
2
\end{array}
\bigg)\frac{1}{M^2}+\bigg(
\begin{array}{c}
n \\
3
\end{array}
\bigg)\frac{29}{12M^3}-\bigg(
\begin{array}{c}
n \\
4
\end{array}
\bigg)\frac{13}{2M^n}+\cdots.
\label{16}
\end{equation}
\begin{figure}[h]
\begin{center}
\includegraphics[scale=0.65]{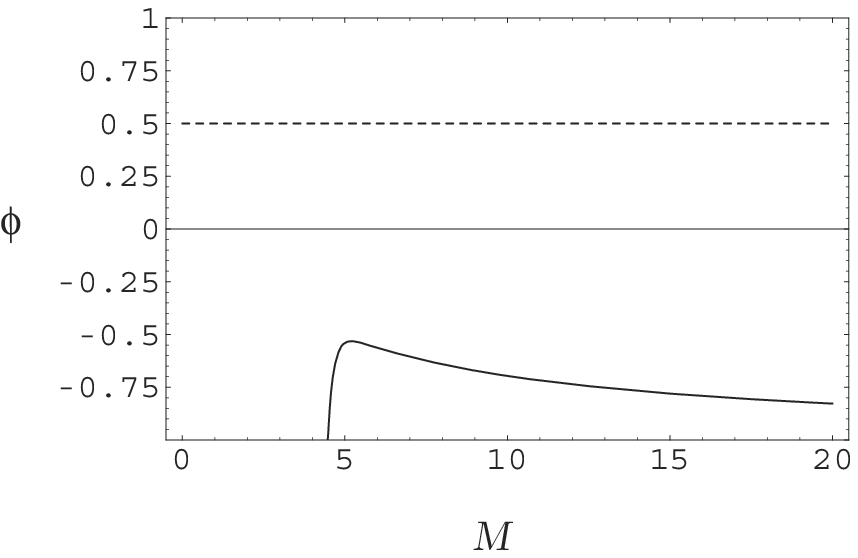}
\includegraphics[scale=0.65]{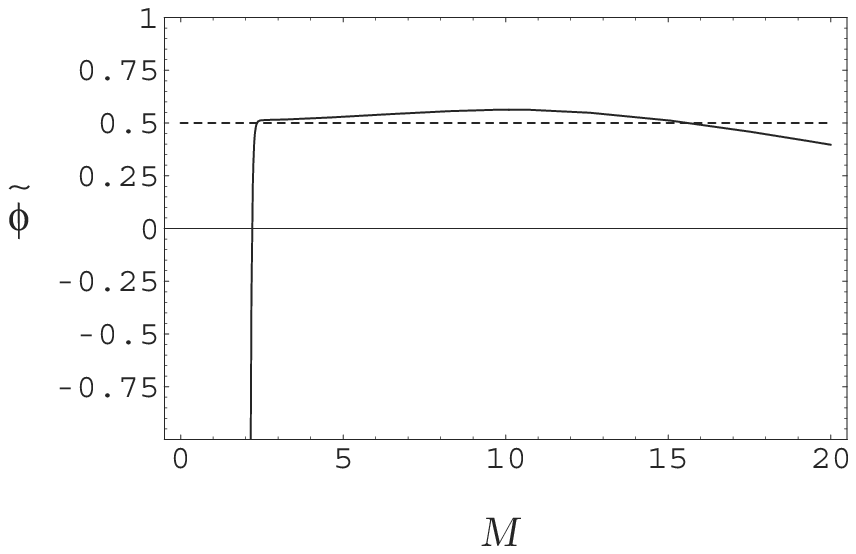}
\caption{Plots of $\phi(M)$ and $\tilde\phi(M)$ as functions of $M$ at $25^{\rm th}$ order.  The dotted lines represent $\frac{1}{2\nu}(=\frac{1}{2})$.}
\end{center}
\end{figure}  
FIG. 9 shows the plots of $\phi(M)$ and $\tilde\phi(M)$ in $1/M$ expansion at $n=25$.  The original function $\phi$ plotted in the left side is far from the scaling region.  On the other hand, $\delta$-expanded series $\tilde\phi$  plotted in the right side shows greatly improved behavior.   First we find that, apart from $\tilde\beta(M)$ where the effects of the corrections to the leading constant $\beta_{c}$ is reduced but non-negligible, the corrections to $\frac{1}{2\nu}$ in $\tilde\phi$ is surpressed enough by delta expansion and the stationary behavior is observed.   
We find that the scaling roughly starts about $M\sim 10$ and abruptly ends around $M\sim 2.5$ which signals the lower end of the region where $\tilde\phi$ in $1/M$ expansion is valid.  We can say that the asymptotic realm just starts at $M\sim 5$ and $\tilde\beta_{n}$ in $1/M$ expansion at $n=25$ surely shows the expected behavior and strongly suggests the correct value of $\nu$.  It is shown that the scaling region is enlarged as to be captured in $1/M$ expansion of $\tilde\beta$.

\section{Conclusion}
Under dilation of the scaling region by shifting $M\to (1-\delta)M$ and the associated expansion in $\delta$, we have shown that $1/M$ expansion effective at large lattice spacing recovered the asymptotic scaling behavior in the non-linear $\sigma$ model at $N=\infty$.  In the large $N$ case, Symanzik's improved action played an important role in the enhancement of the scaling. 

At $N=1$, the scaling behavior was confirmed and rough estimation of the critical exponent $\nu$ is possible.  However, the precision is not so high even at $25^{\rm th}$ order.  Symanzik's improvement scheme would be an promissing candidate to raise the accuracy since, as shown in the $N=\infty$ case , the scheme would reduce the corrections to the asymptotic scaling.

The present work stimulates us to investigate similar subjects on the lattice related to the approximation of the continuum field theories.  Especially, it is of interest to apply the dilation with the delta expansion to lattice  non-Abelian gauge theories.  We hope to report the results in the near future.

\end{document}